\documentclass[12pt]{article}
\begin{document}

\title{Superparticle Sum Rules\\ 
in the presence of\\
Hidden Sector Dynamics}

\author{
Yoshiharu \textsc{Kawamura}\footnote{E-mail: haru@azusa.shinshu-u.ac.jp}, 
Teppei \textsc{Kinami} and Takashi \textsc{Miura}\\
{\it Department of Physics, Shinshu University,}\\
{\it Matsumoto 390-8621, Japan}
}

\date{%
October 22, 2008}

\maketitle
\begin{abstract}
We derive sum rules among scalar masses for various 
boundary conditions of the hidden-visible couplings
in the presence of hidden sector dynamics and
show that they still can be useful probes of the MSSM and beyond.
\end{abstract}

\section{Introduction}

The supersymmetric (SUSY) extension of the standard model has been attractive
as physics beyond the weak scale \cite{N,H&K}. 
The gauge coupling unification can be realized within the framework of the minimal
supersymmetric standard model (MSSM), under the assumption of $\lq$desert' between
the TeV scale and the unification scale \cite{unif,unif2,unif3,unif4}.
It is natural to expect that a similar unification occurs for soft SUSY breaking parameters
at some high-energy scale, reflecting a physics beyond the MSSM \cite{FHK&N,M&R,KM&Y,KM&Y2}.

It is, however, pointed out that hidden sector interactions can give rise to sizable
effects on renormalization group (RG) evolutions of soft SUSY breaking parameters
and some modifications of ordinary analysis are necessary \cite{DFGSS&T}.\footnote{
Conformal sequestering and its phenomenological implications were studied in Refs. \cite{S,IINSY,SS,KMS,MNP}.}
Cohen et al. have derived sfermion mass relations at the TeV scale 
in the presence of hidden sector dynamics, under the assumption that 
a coupling between the MSSM chiral fields 
and hidden vector superfield operators at a unification scale are universal
and the hidden sector is not within the conformal regime \cite{CR&S}.
It is important to examine whether sfermion masses can be useful probes for a high-energy physics,
in the case that the coupling universality is relaxed 
with SUSY grand unified theories (GUTs) in mind.

In this paper, we derive sum rules among scalar masses
for various boundary conditions of the hidden-visible couplings
in the presence of hidden sector effects outside the conformal regime.
We show that their sum rules still can be useful probes of the MSSM and beyond.

The contents of this paper are as follows.
In section 2, we study a modification of RG evolution for scalar masses
by the hidden sector interactions.
In section 3, specific sum rules among scalar masses are derived for various 
boundary conditions of the hidden-visible couplings.
In section 4, sum rules among sfermion masses are also studied for orbifold family unification models.
Section 5 is devoted to conclusions and discussions.

\section{Renormalization group evolution of scalar masses}

\subsection{Basic assumptions}

First we list assumptions adopted in our analysis.\\
1. The theory beyond the SM is the MSSM.
Here the MSSM means the SUSY extension of the SM with the minimal particle contents,
without specifying the structure of soft SUSY breaking terms.
The superpartners and Higgs bosons have a mass whose magnitude is, at most, of order TeV scale.
We neglect the threshold correction at the TeV scale due to the mass difference among the MSSM particles.
Further the TeV scale is often identified with the weak scale ($M_{EW}$) for simplicity.\\
2. The MSSM holds from TeV scale to a high energy scale ($M$).
Above $M$, there is a new physics.
Possible candidates are supergravity (SUGRA), SUSY GUT and/or SUSY orbifold GUT.
There is a big desert between $M_{EW}$ and $M$ in our visible sector.\\
3. The SUSY is broken in a hidden sector at the intermediate scale ($M_I$) 
and the effect is mediated to the visible sector as the appearance of soft SUSY breaking terms.
The hidden sector fields are dynamical from $M$ to $M_I$.
The pattern of soft SUSY breaking parameters reflects on symmetries,
the mechanism of SUSY breaking and the way of its mediation. 
We do not specify the mechanism of SUSY breaking.
In most cases, we assume that the gravity mediation is dominant and
soft SUSY breaking terms respect the gauge invariance.
After the breakdown of gauge symmetry, there appear extra contributions to
soft SUSY breaking parameters, which do not respect the gauge symmetry any more, 
e.g., $D$-term contributions \cite{Dterm,Dterm2,KM&Y,KM&Y2}.
In most case, we consider only $D$-contribution for the electroweak symmetry breaking for simplicity.\\ 
4. The pattern of Yukawa couplings reflects flavor structure in a high-energy theory.
We assume that a suitable pattern of Yukawa couplings is obtained in the low-energy effective theory.
We neglect effects of Yukawa couplings concerning to the first two generations and those of the off-diagonal ones
because they are small compared with the third generation ones.\\
5. The sufficient suppression of flavor-changing neutral currents (FCNC) processes 
requires the mass degeneracy for each squark and slepton species 
in the first two generations unless those masses are rather heavy 
or fermion and its superpartner mass matrices are aligned.
We assume that the generation-changing entries in the sfermion mass matrices are sufficiently small
in the basis where fermion mass matrices are diagonal.
At first, we derive sum rules without the requirement of mass degeneracy and after that we give a brief comment on
the case with the degenerate masses.\\
6. After some parameters are made real by the rephasing of fields, CP violation occurs if the rest are complex.
We assume that Yukawa couplings are dominant as a source of CP violation
and other parameters are real.

\subsection{Renormalization group evolutiuon}

We study RG evolution of scalar mass parameters 
in the presence of hidden sector dynamics \cite{DFGSS&T,CR&S}.
The general hidden sector fields are given by chiral superfield operators $X_x$ and 
vector superfield operators $V_v$ whose auxiliary components are $F_x$ and $D_v$, respectively.
Those fields are treated as dynamical down to the intermediate scale $M_I$.
Visible sector fields consists of chiral superfields $\Phi_{\tilde{F}}$ and spinor superfields $W_i$
whose lowest components are scalar fields $\tilde{F}$ and the MSSM gauginos $\lambda_i$ $(i=1,2,3)$, respectively.
The $\tilde{F}$ represents a multiplet of $G_{SM}=SU(3)_C \times SU(2)_L \times U(1)_Y$, which contains
the scalar partner of the SM fermions and two Higgs doublets $h_{1}$ and $h_{2}$, and are written by,
\begin{eqnarray}
\tilde{F} = \left\{
\begin{array}{ccccc}
\tilde{q}_{1},& \tilde{u}_{R}^*,&\tilde{d}_{R}^*,&\tilde{l}_{1},&\tilde{e}_{R}^*,\\
\tilde{q}_{2},& \tilde{c}_{R}^*,&\tilde{s}_{R}^*,&\tilde{l}_{2},&\tilde{\mu}_{R}^*,\\
\tilde{q}_{3},& \tilde{t}_{R}^*,&\tilde{b}_{R}^*,&\tilde{l}_{3},&\tilde{\tau}_{R}^* ,\\
h_{1},& h_{2},& & & \\
\end{array}
\right. 
\label{F}
\end{eqnarray}
where $\tilde{q}_1$ means the first generation scalar quark (squark) doublet, 
$\tilde{u}^*_R$ up squark singlet, $\tilde{d}^*_R$ down squark singlet,
$\tilde{l}_1$ the first generation scalar lepton (slepton) doublet, $\tilde{e}^*_R$ selectron singlet and so on.
The astrisk means its complex conjugate.

The hidden-visible couplings are given by
\begin{eqnarray}
&~& \sum_{\tilde{F}} \int d^4\theta \sum_v k_{\tilde{F}}^{(v)} \frac{V_v}{M^2} 
\Phi_{\tilde{F}}^{\dagger} \Phi_{\tilde{F}}
\nonumber \\
&~& ~~~ + \sum_i \int d^2\theta \sum_x w_i^{(x)} \frac{X_x}{M} W_i W_i + \mbox{h.c.} 
\nonumber \\
&~& ~~~ + \sum_{r}  
\int d^2\theta \sum_x a_r^{(x)} f_r \frac{X_x}{M} \Phi_{\tilde{F}} \Phi_{\tilde{F}'} \Phi_{\tilde{F}''}+ \mbox{h.c.} 
\nonumber \\
&~& ~~~ + 
\int d^2\theta \sum_x b^{(x)} \mu \frac{X_x}{M} H_1 H_2 + \mbox{h.c.} ,
\label{hv-coupling}
\end{eqnarray}
where h.c. means the hermitian conjugate of the former term
and $r$ represents indices regarding trilinear couplings (and Yukawa couplings) among visible sector fields, e.g.,
\begin{eqnarray}
r = \left\{
\begin{array}{cc}
t & \mbox{for} ~~ (\tilde{q}_{3}, \tilde{t}_R^*, h_2) , \\
b & \mbox{for} ~~ (\tilde{q}_{3}, \tilde{b}_R^*, h_1) , \\
\tau & \mbox{for} ~~ (\tilde{l}_{3}, \tilde{\tau}_R^*, h_1) .
\end{array}
\right. 
\label{r}
\end{eqnarray}
In (\ref{hv-coupling}), we assume that there is no flavor mixing in the first term 
and trilinear couplings among visible sector fields exist only in the third generation.
Scalar mass-squareds $m_{\tilde{F}}^2$, gaugino masses $M_i$, $A$-parameters and $B$-parameter are given by
\begin{eqnarray}
&~& m_{\tilde{F}}^2(t_I) = \sum_v k_{\tilde{F}}^{(v)}(t_I) \frac{\langle D_v \rangle}{M^2} ,
\label{mF}\\
&~& M_i(t_I) = \sum_x w_{i}^{(x)}(t_I) \frac{\langle F_x \rangle}{M} g_i^2(t_I) ,
\label{Mi}\\
&~& A_r(t_I) = \sum_x a_r^{(x)}(t_I) \frac{\langle F_x \rangle}{M} ,
\label{A}\\
&~& B(t_I) = \sum_x b^{(x)}(t_I) \frac{\langle F_x \rangle}{M} ,
\label{B}
\end{eqnarray}
where $\displaystyle{t_I \equiv \frac{1}{2\pi} \ln(M/M_I)}$ and
$g_i$s are gauge couplings of $G_{\rm SM}$. 
The RG equation regarding $k_{\tilde{F}}^{(v)}$ is given by
\begin{eqnarray}
&~& \frac{{\rm d}\ }{{\rm d}t}k_{\tilde{F}}^{(v)} =
  - \sum_{v'} \gamma_{vv'} k_{\tilde{F}}^{(v')} + \frac{1}{8\pi} \sum_i 8 C_2^{(i)}(\tilde{F}) g_i^6 G_i^{(v)}
\nonumber \\
&~& ~~~~~ - \frac{1}{4\pi} \frac{3}{5} Y(\tilde{F}) g_1^2 k_S^{(v)} 
  - \frac{1}{4\pi} \sum_r n_{\tilde{F}}^{(r)} f_r^2 \left(k_r^{(v)} + h_r^{(v)}\right) ,
\label{k-RGE}
\end{eqnarray}
where $\displaystyle{t \equiv \frac{1}{2\pi} \ln(M/\mu)}$ and $\mu$ is the renormalization scale.
The $\gamma_{vv'}$ is the anomalous dimension matrix of $V_v$.
The $C_2^{(i)}(\tilde{F})$ and $Y(\tilde{F})$ 
represent the eigenvalues of second Casimir operator 
(e.g., $C_2^{(3)}({\tilde{q}_1})=4/3$, $C_2^{(2)}({\tilde{q}_1})=3/4$ and $C_2^{(1)}({\tilde{q}_1})=1/60$)
and hypercharge for $\tilde{F}$, respectively. 
The $n_{\tilde{F}}^{(r)}$ are given by
\begin{eqnarray}
&~& n_{\tilde{t}_L}^{(t)} = n_{\tilde{t}_L}^{(b)} = n_{\tilde{b}_L}^{(t)} = n_{\tilde{b}_L}^{(b)} 
= n_{\tilde{\nu}_{\tau L}}^{(\tau)} = n_{\tilde{\tau}_{L}}^{(\tau)} = n_{h_1}^{(\tau)} = 1 ,
\nonumber \\
&~& n_{\tilde{t}_R^*}^{(t)} = n_{\tilde{b}_R^*}^{(b)} = n_{\tilde{\tau}_{R}^*}^{(\tau)} = 2 , ~~
n_{h_1}^{(b)} = n_{h_2}^{(t)} = 3 .
\label{a(f)}
\end{eqnarray}
The $G_i^{(v)}$, $k_S^{(v)}$, $k_r^{(v)}$ and $h_r^{(v)}$ are defined by
\begin{eqnarray}
&~& G_i^{(v)} \equiv \sum_{x, x'} w_i^{*(x)} J_{xx'}^{(v)} w_i^{(x')} , ~~
k_S^{(v)} \equiv \sum_{\tilde{F}} Y(\tilde{F}) n_{\tilde{F}} k_{\tilde{F}}^{(v)} ,
\label{GkS}\\
&~& k_t^{(v)} \equiv k_{\tilde{q}_3}^{(v)} + k_{\tilde{t}_R^*}^{(v)} + k_{h_2}^{(v)} , ~~
k_b^{(v)} \equiv k_{\tilde{q}_3}^{(v)} + k_{\tilde{b}_R^*}^{(v)} + k_{h_1}^{(v)} , 
\nonumber \\
&~& k_{\tau}^{(v)} \equiv k_{\tilde{l}_3}^{(v)} + k_{\tilde{\tau}_R^*}^{(v)} + k_{h_1}^{(v)} ,
\label{kf}\\
&~& h_r^{(v)} \equiv \sum_{x, x'} a_r^{*(x)} J_{xx'}^{(v)} a_r^{(x')} ,
\end{eqnarray}
where $J_{xx'}^{(v)}$ stands for a factor from the interaction among $X_x$, $X_{x'}$ and $V_v$, 
and $n_{\tilde{F}}$ represents degrees of freedom for $\tilde{F}$.
The $k_S^{(v)}$ yields the following RG equations,
\begin{eqnarray}
&~& \frac{{\rm d}\ }{{\rm d}t}k_{S}^{(v)} = - \sum_{v'} 
\left(\gamma_{vv'} + b_1 \alpha_1 \delta_{vv'}\right) k_{S}^{(v')} .
\label{kS-RGE}
\end{eqnarray}
By integrating (\ref{k-RGE}) and (\ref{kS-RGE}), we obtain the following expressions for $k_{\tilde{F}}(t)$
and $k_S(t)$:
\begin{eqnarray}
\hspace{-14mm} &~& k_{\tilde{F}}(t) = {\rm P}\exp\left(-\int_0^t dt' \gamma(t')\right) k_{\tilde{F}}(0)
\nonumber \\
\hspace{-14mm} &~& ~~~~~  +  \frac{1}{8\pi} \sum_i 8 C_2^{(i)}(\tilde{F}) \int_0^t ds 
 {\rm P}\exp\left(-\int_s^t dt' \gamma(t')\right) g_i^6(s) G_i(s)
\nonumber \\
\hspace{-14mm} &~& ~~~~~ - \frac{1}{4\pi} \frac{3}{5} Y(\tilde{F}) \int_0^t ds
 {\rm P}\exp\left(-\int_s^t dt' \gamma(t')\right)  g_1^2(s) k_S(s) 
\nonumber \\
\hspace{-14mm} &~& ~~~~~  - \frac{1}{4\pi} \sum_r n_{\tilde{F}}^{(r)} \int_0^t ds 
 {\rm P}\exp\left(-\int_s^t dt' \gamma(t')\right) f_r^2(s) \left(k_r(s) + h_r(s)\right) ,
\label{k-sol}\\
\hspace{-14mm} &~& k_{S}(t) = {\rm P}\exp\left(-\int_0^t dt' \left(\gamma(t') + b_1 \alpha_1(t')\right)\right) k_{S}(0) ,
\label{kS-sol}
\end{eqnarray}
where the index $v$ is suppressed and ${\rm P}$ represents the path-ordered exponentials.
Scalar mass-squareds $m_{\tilde{F}}^2(t_I)$ are obtained by inserting (\ref{k-sol}) into the formula (\ref{mF}).
Further the $m_{\tilde{F}}^2$ at the TeV scale ($M_{EW}$) are written by
\begin{eqnarray}
&~& m_{\tilde{F}}^{2}(t_{EW}) = m_{\tilde{F}}^{2}(t_I) 
 + \sum_{i=1}^{3}\frac{2 C_2^{(i)}(\tilde{F})}{b_{i}}\left(M_{i}^{2}(t_I)-M_{i}^{2}(t_{EW})\right) 
\nonumber\\
&~& ~~~~~~~~~~~~~~~ +\frac{3}{5b_{1}}Y(\tilde{F})\left(S(t_{EW})-S(t_I)\right)  
 + \sum_r n_{\tilde{F}}^{(r)} \left(F_r(t_{EW})-F_r(t_I)\right)  
\nonumber \\
&~&  ~~~~~~~~~~~~ = N_{\tilde{F}} + \sum_{i=1}^{3} C_2^{(i)}(\tilde{F}) N_i + Y(\tilde{F}) N_S 
 + \sum_r n_{\tilde{F}}^{(r)} N_r ,
\label{mF(t)}
\end{eqnarray}
where $\displaystyle{t_{EW} \equiv \frac{1}{2\pi} \ln(M_I/M_{EW})}$.
In the final expression, $N_{\tilde{F}}$, $N_i$, $N_S$ and $N_r$ are defined by
\begin{eqnarray}
&~& N_{\tilde{F}} \equiv \sum_v \frac{\langle D \rangle}{M^2} 
 {\rm P}\exp\left(-\int_0^{t_I} dt' \gamma(t')\right) k_{\tilde{F}}(0) ,
\label{NF}\\
&~& N_i \equiv \frac{1}{\pi} \sum_v \frac{\langle D \rangle}{M^2} 
 \int_0^{t_I} ds {\rm P}\exp\left(-\int_s^{t_I} dt' \gamma(t')\right) g_i^6(s)  G_i(s) 
\nonumber \\
&~& ~~~~~~~ + \frac{2}{b_{i}}\left(M_{i}^{2}(t_I)-M_{i}^{2}(t_{EW})\right) ,
\label{Ni}\\
&~& N_S \equiv - \frac{1}{4\pi} \frac{3}{5} \sum_v \frac{\langle D \rangle}{M^2} 
 \int_0^{t_I} ds {\rm P}\exp\left(-\int_s^{t_I} dt' \gamma(t')\right) g_1^2(s)  k_S(s) 
\nonumber \\
&~& ~~~~~~~ + \frac{3}{5b_{1}}\left(S(t_{EW})-S(t_I)\right) ,
\label{NS}\\
&~& N_r \equiv - \frac{1}{4\pi} \sum_v \frac{\langle D \rangle}{M^2}
\int_0^{t_I} ds {\rm P}\exp\left(-\int_s^{t_I} dt' \gamma(t')\right) f_r^2(s) \left(k_r(s) + h_r(s)\right)  
\nonumber \\
&~& ~~~~~~~ + F_r(t_{EW})-F_r(t_I) ,
\label{Nr} 
\end{eqnarray}
where we use the conventional RG equations in the MSSM from $t_I$ to $t_{EW}$ such that\cite{Kazakov,Polonsky,DG&R,B&T}
\begin{eqnarray}
&~& \frac{{\rm d}\ }{{\rm d}t}m_{\tilde{F}}^{2} = 4 \sum_{i=1}^{3}C_2^{(i)}(\tilde{F}) \alpha_{i}M_{i}^{2}
 -\frac{3}{5}Y(\tilde{F})\alpha_{1}S 
\nonumber \\
&~& ~~~~~~~~~~~~ - \sum_r n_{\tilde{F}}^{(r)} \frac{f_r^2}{4\pi} \left({\sum}'_{\tilde{F}} m_{\tilde{F}}^2 + A_r^2\right) ,
\label{mF-RGE}\\
&~& \frac{{\rm d}\ }{{\rm d}t}S = -b_{1}\alpha_{1}S ,~~ S \equiv \sum_{\tilde{F}}Y(\tilde{F})n_{\tilde{F}}m_{\tilde{F}}^{2} ,
\label{S-RGE}
\end{eqnarray}
where ${\sum}'_{\tilde{F}}$ means a sum among scalar masses relating to Yukawa interactions.
The $F_r$s in (\ref{mF(t)}) and (\ref{Nr}) stand for contributions from Yukawa interactions and satisfy the following equations
\begin{eqnarray}
&~& \frac{{\rm d}\ }{{\rm d}t}F_t = \frac{f_t^2}{4\pi}\left(m_{\tilde{q}_3}^2 + m_{\tilde{t}_R}^2 + m_{h_2}^2 + A_t^2\right) , 
\label{Ft} \\
&~& \frac{{\rm d}\ }{{\rm d}t}F_b = \frac{f_b^2}{4\pi}\left(m_{\tilde{q}_3}^2 + m_{\tilde{b}_R}^2 + m_{h_1}^2 + A_b^2\right) , 
\label{Fb} \\
&~& \frac{{\rm d}\ }{{\rm d}t}F_{\tau} 
= \frac{f_{\tau}^2}{4\pi}\left(m_{\tilde{l}_3}^2 + m_{\tilde{\tau}_R}^2 + m_{h_1}^2 + A_{\tau}^2\right) . 
\label{Ftau} 
\end{eqnarray}
Complete analytic solutions for $F_t$, $F_b$ and $F_{\tau}$ are not known and 
those values are determined numerically by solving RG equations of sparticle masses and coupling constants.
We treat $N_{\tilde{F}}$, $N_i$, $N_S$ and $N_r$ as free parameters
because $\gamma(t')$ is a unknown function, which reflects on the hidden sector dynamics.

After the breakdown of electroweak symmetry, two kinds of contributions are added to sfermion masses, i.e.,
fermion masses ($m_f$) and the $D$-term contribution ($D_{W}(\tilde{f})$) relating to 
the generator of the broken symmetry ($SU(2)_{L} \times U(1)_{Y})/U(1)_{EM}$.
The diagonal elements ($M_{\tilde{f}}^2$) of sfermion mass-squared matrices at $M_{EW}$ are written as
\begin{eqnarray}
\hspace{-10mm} &~& M_{\tilde{f}}^{2} = m_{\tilde{F}}^{2} + m_{f}^{2} + D_{W}(\tilde{f}) ,
\nonumber \\
\hspace{-10mm} &~& ~~~~ = N_{\tilde{F}} + \sum_{i=1}^{3} C_2^{(i)}(\tilde{F}) N_i + Y(\tilde{F}) N_S 
 + \sum_r n_{\tilde{F}}^{(r)} N_r + m_{f}^{2} + D_{W}(\tilde{f}) .
\label{Mf}
\end{eqnarray}
where $\tilde{f}$ means the scalar partner of fermion species $f$.
The $f$s are given by 
\begin{eqnarray}
f = \left\{
\begin{array}{ccccccc}
{u}_{L},& {d}_{L},& {u}_{R},& {d}_{R},& {\nu}_{eL},& {e}_{L},& {e}_{R},\\
{c}_{L},& {s}_{L},& {c}_{R},& {s}_{R},& {\nu}_{\mu L},& {\mu}_{L},& {\mu}_{R},\\
{t}_{L},& {b}_{L},& {t}_{R},& {b}_{R},& {\nu}_{\tau L},& {\tau}_{L},& {\tau}_{R} .
\end{array}
\right. 
\label{f}
\end{eqnarray}
The $D_{W}(\tilde{f})$ are given by
\begin{eqnarray}
\hspace{-10mm} &~& D_{W}(\tilde{f}) = \left(T_{L}^{3}(\tilde{f})-Q(\tilde{f})\sin^{2}\theta_{W}\right) M_{Z}^{2}\cos 2\beta 
\nonumber \\
\hspace{-10mm} &~& ~~~~~~~~~ = \left(\left(T_{L}^{3}-Q(\tilde{f})\right)M_{Z}^{2}+Q(\tilde{f})M_{W}^{2}\right)
 \cos 2\beta ~~ (f = u_L, \cdots \tau_L) ,
\label{DfL}\\
\hspace{-10mm} &~& D_{W}(\tilde{f}) =  Q(\tilde{f})\sin^{2}\theta_{W} M_{Z}^{2}\cos 2\beta 
\nonumber \\
\hspace{-10mm} &~& ~~~~~~~~~~ = Q(\tilde{f})\left(M_{Z}^{2} - M_{W}^{2}\right)\cos 2\beta ~~~~~~~~~~~~~~~ (f = u_R, \cdots \tau_R) .
\label{DfR}
\end{eqnarray}
The off-diagonal elements of sfermion mass-squared matrices are proportional to the corresponding fermion mass.
For  the first two generations, the diagonal ones $M_{\tilde{f}}^{2}$ are regarded as $\lq$physical masses' 
which are eigenvalues of mass-squared matrices because the off-diagonal ones are negligibly small.
Using the mass formula (\ref{Mf}), values of ${m}_{\tilde{F}}^{2}$ can be determined for the first two generations.
For the third generation, mass-squared matrices are given by
\begin{eqnarray}
\hspace{-13mm} &~& \left(
\begin{array}{cc}
 m_{\tilde{t}_L}^{2} + m_{t}^{2} + D_{W}(\tilde{t}_L) & -m_t(A_t + \mu \cot\beta) \\
 -m_t(A_t + \mu \cot\beta) & m_{\tilde{t}_R}^{2} + m_{t}^{2} + D_{W}(\tilde{t}_R) 
\end{array}
\right) ~~ (\mbox{for top squarks}),
\label{mass-stop}\\
\hspace{-13mm} &~& \left(
\begin{array}{cc}
 m_{\tilde{b}_L}^{2} + m_{b}^{2} + D_{W}(\tilde{b}_L) & -m_b(A_b + \mu \tan\beta) \\
 -m_b(A_b + \mu \tan\beta) & m_{\tilde{b}_R}^{2} + m_{b}^{2} + D_{W}(\tilde{b}_R) 
\end{array}
\right)  (\mbox{for bottom squarks}),
\label{mass-sbottom}\\
\hspace{-13mm} &~& \left(
\begin{array}{cc}
 m_{\tilde{\tau}_L}^{2} + m_{\tau}^{2} + D_{W}(\tilde{\tau}_L) & -m_{\tau}(A_{\tau} + \mu \tan\beta) \\
 -m_{\tau}(A_{\tau} + \mu \tan\beta) & m_{\tilde{\tau}_R}^{2} + m_{\tau}^{2} + D_{W}(\tilde{\tau}_R) 
\end{array}
\right) ~~ (\mbox{for tau sleptons}) .
\label{mass-stau}
\end{eqnarray}
By diagonalized the above mass-squared matrices, we obtain mass eigenstates whose masses are physical ones,
($M_{\tilde{t}_1}$, $M_{\tilde{t}_2}$) for top squarks, ($M_{\tilde{b}_1}$, $M_{\tilde{b}_2}$) for bottom squarks
and ($M_{\tilde{\tau}_1}$, $M_{\tilde{\tau}_2}$) for tau sleptons.
By using the feature of trace, we have the relations,
\begin{eqnarray}
&~& M_{\tilde{t}_1}^2 + M_{\tilde{t}_2}^2 = M_{\tilde{t}_L}^2 + M_{\tilde{t}_R}^2 , ~~
M_{\tilde{b}_1}^2 + M_{\tilde{b}_2}^2 = M_{\tilde{b}_L}^2 + M_{\tilde{b}_R}^2 ,
\nonumber \\
&~& M_{\tilde{\tau}_1}^2 + M_{\tilde{\tau}_2}^2 = M_{\tilde{\tau}_L}^2 + M_{\tilde{\tau}_R}^2 .
\label{trace}
\end{eqnarray}
By diagonalizing the mass squared matrices, we have the relations,
\begin{eqnarray}
&~& \left(M_{\tilde{t}_1}^2 - M_{\tilde{t}_2}^2\right)^2 = 
 \left(M_{\tilde{t}_L}^2 - M_{\tilde{t}_R}^2\right)^2 + 4 m_t^2 \left(A_t + \mu \cot\beta\right)^2 , ~~
\label{t-}\\
&~& \left(M_{\tilde{b}_1}^2 - M_{\tilde{b}_2}^2\right)^2 = 
 \left(M_{\tilde{b}_L}^2 - M_{\tilde{b}_R}^2\right)^2 + 4 m_b^2 \left(A_b + \mu \tan\beta\right)^2 , ~~
\label{b-}\\
&~& \left(M_{\tilde{\tau}_1}^2 - M_{\tilde{\tau}_2}^2\right)^2 = 
 \left(M_{\tilde{\tau}_L}^2 - M_{\tilde{\tau}_R}^2\right)^2 + 4 m_{\tau}^2 \left(A_{\tau} + \mu \tan\beta\right)^2 , ~~
\label{t-}
\end{eqnarray}
If $A$ parameters were measured precisely,
${m}_{\tilde{F}}^{2}$s (and ${M}_{\tilde{f}}^{2}$s) in the third generation can be fixed 
by using the mass-squared matrices (\ref{mass-stop}) - (\ref{mass-stau}).
{}From the fact that left-handed fermions (and its superpartners) form $SU(2)_L$ doublets, e.g., $q_1 = (u_L, d_L)$ 
(and $\tilde{q}_1 = (\tilde{u}_L, \tilde{d}_L)$),
we obtain following sum rules among $SU(2)_L$ doublet sfermions:
\begin{eqnarray}
&~& M_{\tilde{u}_L}^2 - M_{\tilde{d}_L}^2 = m_u^2 - m_d^2 + M_W^2 \cos2\beta \simeq  M_W^2 \cos2\beta ,
\label{u-d} \\
&~& M_{\tilde{\nu}_{eL}}^2 - M_{\tilde{e}_L}^2 = m_{\nu_{eL}}^2 - m_e^2 + M_W^2 \cos2\beta \simeq  M_W^2 \cos2\beta ,
\label{nu-e} \\
&~& M_{\tilde{c}_L}^2 - M_{\tilde{s}_L}^2 = m_c^2 - m_s^2 + M_W^2 \cos2\beta \simeq  M_W^2 \cos2\beta ,
\label{c-s} \\
&~& M_{\tilde{\nu}_{\mu L}}^2 - M_{\tilde{\mu}_L}^2 = m_{\nu_{\mu L}}^2 - m_\mu^2 + M_W^2 \cos2\beta \simeq  M_W^2 \cos2\beta ,
\label{nu-mu} \\
&~& M_{\tilde{t}_L}^2 - M_{\tilde{b}_L}^2 = m_t^2 - m_b^2 + M_W^2 \cos2\beta \simeq  m_t^2 + M_W^2 \cos2\beta ,
\label{t-b} \\
&~& M_{\tilde{\nu}_{\tau L}}^2 - M_{\tilde{\tau}_L}^2 = m_{\nu_{\tau L}}^2 - m_\tau^2 + M_W^2 \cos2\beta \simeq  M_W^2 \cos2\beta ,
\label{nu-tau} 
\end{eqnarray}
where we neglect fermion masses except for the top quark mass in the final expressions.
The above sum rules (\ref{u-d}) - (\ref{nu-tau}) are irrelevant to the structure of models beyond the MSSM, and hence
the sfermion sector (and the breakdown of electroweak symmetry) in the MSSM can be tested by using them.
We refer to these sum rules (\ref{u-d}) - (\ref{nu-tau}) as the electroweak symmetry (EWS) sum rules.

\section{Sparticle sum rules}

First of all, we write down the scalar masses at $M_{EW}$ using the mass formula (\ref{Mf}), 
\begin{eqnarray}
\hspace{-12mm} &~& M_{\tilde{u}_{L}}^{2} 
= N_{\tilde{q}_1} + \frac{4}{3} N_3 + \frac{3}{4} N_2 + \frac{1}{60} N_1 + \frac{1}{6} N_S 
+ \left(\frac{2}{3}M_{W}^{2} - \frac{1}{6}M_{Z}^{2}\right)\cos 2\beta , \\
\hspace{-12mm} &~& M_{\tilde{d}_{L}}^{2}
= N_{\tilde{q}_1} + \frac{4}{3} N_3 + \frac{3}{4} N_2 + \frac{1}{60} N_1 + \frac{1}{6} N_S 
+ \left(-\frac{1}{3}M_{W}^{2} - \frac{1}{6}M_{Z}^{2}\right)\cos 2\beta , \\
\hspace{-12mm} &~& M_{\tilde{u}_{R}}^{2} 
= N_{\tilde{u}^*_R} + \frac{4}{3} N_3 + \frac{4}{15} N_1 - \frac{2}{3} N_S 
+ \left(-\frac{2}{3}M_{W}^{2}+\frac{2}{3}M_{Z}^{2}\right)\cos 2\beta , \\
\hspace{-12mm} &~& M_{\tilde{d}_{R}}^{2} 
= N_{\tilde{d}^*_R} + \frac{4}{3} N_3 + \frac{1}{15} N_1 + \frac{1}{3} N_S 
+\left(\frac{1}{3}M_{W}^{2} -\frac{1}{3}M_{Z}^{2}\right)\cos 2\beta , \\
\hspace{-12mm} &~& M_{\tilde{\nu}_{eL}}^{2} 
= N_{\tilde{l}_1} + \frac{3}{4} N_2 + \frac{3}{20} N_1 - \frac{1}{2} N_S 
+\frac{1}{2}M_{Z}^{2}\cos 2\beta , \\
\hspace{-12mm} &~& M_{\tilde{e}_{L}}^{2} 
= N_{\tilde{l}_1} + \frac{3}{4} N_2 + \frac{3}{20} N_1 - \frac{1}{2} N_S 
+\left(-M_{W}^{2}+\frac{1}{2}M_{Z}^{2}\right)\cos 2\beta , \\
\hspace{-12mm} &~& M_{\tilde{e}_{R}}^{2} 
= N_{\tilde{e}^*_R} + \frac{3}{5} N_1 + N_S 
+\left(M_{W}^{2}-M_{Z}^{2}\right)\cos 2\beta , \\
\hspace{-12mm} &~& M_{\tilde{c}_{L}}^{2} 
= N_{\tilde{q}_2} + \frac{4}{3} N_3 + \frac{3}{4} N_2 + \frac{1}{60} N_1 + \frac{1}{6} N_S 
+ \left(\frac{2}{3}M_{W}^{2} - \frac{1}{6}M_{Z}^{2}\right)\cos 2\beta , \\
\hspace{-12mm} &~& M_{\tilde{s}_{L}}^{2}
= N_{\tilde{q}_2} + \frac{4}{3} N_3 + \frac{3}{4} N_2 + \frac{1}{60} N_1 + \frac{1}{6} N_S 
+ \left(-\frac{1}{3}M_{W}^{2} - \frac{1}{6}M_{Z}^{2}\right)\cos 2\beta , \\
\hspace{-12mm} &~& M_{\tilde{c}_{R}}^{2} 
= N_{\tilde{c}^*_R} + \frac{4}{3} N_3 + \frac{4}{15} N_1 - \frac{2}{3} N_S 
+ \left(-\frac{2}{3}M_{W}^{2}+\frac{2}{3}M_{Z}^{2}\right)\cos 2\beta , \\
\hspace{-12mm} &~& M_{\tilde{s}_{R}}^{2} 
= N_{\tilde{s}^*_R} + \frac{4}{3} N_3 + \frac{1}{15} N_1 + \frac{1}{3} N_S 
+\left(\frac{1}{3}M_{W}^{2} -\frac{1}{3}M_{Z}^{2}\right)\cos 2\beta , \\
\hspace{-12mm} &~& M_{\tilde{\nu}_{\mu L}}^{2} 
= N_{\tilde{l}_2} + \frac{3}{4} N_2 + \frac{3}{20} N_1 - \frac{1}{2} N_S 
+\frac{1}{2}M_{Z}^{2}\cos 2\beta , \\
\hspace{-12mm} &~& M_{\tilde{\mu}_{L}}^{2} 
= N_{\tilde{l}_2} + \frac{3}{4} N_2 + \frac{3}{20} N_1 - \frac{1}{2} N_S 
+\left(-M_{W}^{2}+\frac{1}{2}M_{Z}^{2}\right)\cos 2\beta , \\
\hspace{-12mm} &~& M_{\tilde{\mu}_{R}}^{2} 
= N_{\tilde{\mu}^*_R} + \frac{3}{5} N_1 + N_S 
+\left(M_{W}^{2}-M_{Z}^{2}\right)\cos 2\beta , \\
\hspace{-12mm} &~& M_{\tilde{t}_{L}}^{2} 
= N_{\tilde{q}_3} + \frac{4}{3} N_3 + \frac{3}{4} N_2 + \frac{1}{60} N_1 + \frac{1}{6} N_S 
+ \left(\frac{2}{3}M_{W}^{2} - \frac{1}{6}M_{Z}^{2}\right)\cos 2\beta 
\nonumber \\
\hspace{-12mm} &~& ~~~~~~~~ + N_t + N_b + m_t^2 , \\
\hspace{-12mm} &~& M_{\tilde{b}_{L}}^{2}
= N_{\tilde{q}_3} + \frac{4}{3} N_3 + \frac{3}{4} N_2 + \frac{1}{60} N_1 + \frac{1}{6} N_S 
+ \left(-\frac{1}{3}M_{W}^{2} - \frac{1}{6}M_{Z}^{2}\right)\cos 2\beta 
\nonumber \\
\hspace{-12mm} &~& ~~~~~~~~ + N_t + N_b + m_b^2 , \\
\hspace{-12mm} &~& M_{\tilde{t}_{R}}^{2} 
= N_{\tilde{t}^*_R} + \frac{4}{3} N_3 + \frac{4}{15} N_1 - \frac{2}{3} N_S 
+ \left(-\frac{2}{3}M_{W}^{2}+\frac{2}{3}M_{Z}^{2}\right)\cos 2\beta  
\nonumber \\
\hspace{-12mm} &~& ~~~~~~~~ + 2 N_t + m_t^2 , \\
\hspace{-12mm} &~& M_{\tilde{b}_{R}}^{2} 
= N_{\tilde{b}^*_R} + \frac{4}{3} N_3 + \frac{1}{15} N_1 + \frac{1}{3} N_S 
+\left(\frac{1}{3}M_{W}^{2} -\frac{1}{3}M_{Z}^{2}\right)\cos 2\beta  
\nonumber \\
\hspace{-12mm} &~& ~~~~~~~~ + 2 N_b + m_b^2 , \\
\hspace{-12mm} &~& M_{\tilde{\nu}_{\tau L}}^{2} 
= N_{\tilde{l}_3} + \frac{3}{4} N_2 + \frac{3}{20} N_1 - \frac{1}{2} N_S 
+\frac{1}{2}M_{Z}^{2}\cos 2\beta  
\nonumber \\
\hspace{-12mm} &~& ~~~~~~~~ + N_{\tau} , \\
\hspace{-12mm} &~& M_{\tilde{\tau}_{L}}^{2} 
= N_{\tilde{l}_3} + \frac{3}{4} N_2 + \frac{3}{20} N_1 - \frac{1}{2} N_S 
+\left(-M_{W}^{2}+\frac{1}{2}M_{Z}^{2}\right)\cos 2\beta 
\nonumber \\
\hspace{-12mm} &~& ~~~~~~~~ + N_{\tau} + m_{\tau}^2 , \\
\hspace{-12mm} &~& M_{\tilde{\tau}_{R}}^{2} 
= N_{\tilde{\tau}^*_R} + \frac{3}{5} N_1 + N_S 
+\left(M_{W}^{2}-M_{Z}^{2}\right)\cos 2\beta 
\nonumber \\
\hspace{-12mm} &~& ~~~~~~~~ + 2 N_{\tau} + m_{\tau}^2 , \\
\hspace{-12mm} &~& m_{h_1}^2 
= N_{h_1} + \frac{3}{4} N_2 + \frac{3}{20} N_1 + \frac{1}{2} N_S + N_{\tau} + 3N_b , \\
\hspace{-12mm} &~& m_{h_2}^2 
= N_{h_2} + \frac{3}{4} N_2 + \frac{3}{20} N_1 - \frac{1}{2} N_S +  3N_t ,
\end{eqnarray}
where we neglect effects of Yukawa couplings in the first two generations.
Extra $D$-term contributions are not written because they depend on a large gauge group.
Hereafter we neglect $m_b$ and $m_{\tau}$ for simplicity.

In the next section, we derive specific sum rules (except for the EWS sum rules)
reflecting on the structure of hidden-visible couplings
for various ultra-violet (UV) boundary conditions

\subsection{Universal type}

Let us discuss the case with a universal hidden-visible coupling at $M$, i.e., $k_{\tilde{F}}^{(v)}(0) = k_0$.
In this case, $N_{\tilde{F}}$ takes a common value and $N_S = 0$.
There exists a specific sum rule among the first generation sfermion masses such as\cite{CR&S}
\begin{eqnarray}
2M_{\tilde{u}_{R}}^{2}-M_{\tilde{d}_{R}}^{2}-M_{\tilde{d}_{L}}^{2}+M_{\tilde{e}_{L}}^{2}-M_{\tilde{e}_{R}}^{2} 
 = \frac{10}{3}\left(M_{Z}^{2}-M_{W}^{2}\right)\cos 2\beta .
\label{IG}
\end{eqnarray}
There exist five kinds of sum rules among first and second generations sfermion masses such that
\begin{eqnarray}
&~& M_{\tilde{u}_{L}}^{2}-M_{\tilde{c}_{L}}^{2} = M_{\tilde{u}_{R}}^{2}-M_{\tilde{c}_{R}}^{2} 
= M_{\tilde{d}_{R}}^{2}-M_{\tilde{s}_{R}}^{2} 
\nonumber \\
&~& ~~~~~ = M_{\tilde{e}_{L}}^{2}-M_{\tilde{\mu}_{L}}^{2} = M_{\tilde{e}_{R}}^{2}-M_{\tilde{\mu}_{R}}^{2} 
= 0 .
\label{12}
\end{eqnarray}
Further we obtain four kinds of sum rules including third generation sfermion masses and/or Higgs masses
such that
\begin{eqnarray}
\hspace{-8mm} &~& 2\left(M_{\tilde{u}_{L}}^{2}-M_{\tilde{t}_{L}}^{2} + m_t^2\right) 
= M_{\tilde{u}_{R}}^{2} + M_{\tilde{d}_{R}}^{2} - M_{\tilde{t}_{R}}^{2} - M_{\tilde{b}_{R}}^{2}  + m_t^2 ,
\label{13-1}\\
\hspace{-8mm} &~& 2\left(M_{\tilde{e}_{L}}^{2}-M_{\tilde{\tau}_{L}}^{2}\right) 
= M_{\tilde{e}_{R}}^{2} - M_{\tilde{\tau}_{R}}^{2} ,
\label{13-2}\\
\hspace{-8mm} &~& 2\left(m_{h_1}^2 - m_{h_2}^2\right) = 2\left(M_{\tilde{\tau}_{L}}^{2}-M_{\tilde{e}_{L}}^{2}\right)
\nonumber \\
\hspace{-8mm} &~& ~~~~~ + 3\left(M_{\tilde{b}_{R}}^{2}-M_{\tilde{d}_{R}}^{2} 
 + M_{\tilde{u}_{R}}^{2} - M_{\tilde{t}_{R}}^{2} + m_{t}^{2}\right) ,
\label{13-3}\\
\hspace{-8mm} &~& 2 \left(m_{h_2}^2 - M_{\tilde{e}_{L}}^{2}\right) = 
  3\left(M_{\tilde{t}_{R}}^{2}-M_{\tilde{u}_{R}}^{2} - m_{t}^{2}\right) 
 + \left(2M_{W}^{2}-M_{Z}^{2}\right)\cos 2\beta .
\label{13-4}
\end{eqnarray}
If all paremeters were measured precisely enough, these sum rules can be powerful tools to test the universality of
$k_{\tilde{F}}^{(v)}$ at $M$.

Here we give two comments for a later convenience.
In the case with a non-vanishing $N_S$, 
the sum rules (\ref{12}), (\ref{13-1}), (\ref{13-2}) and (\ref{13-4}) hold on.\footnote{
If there were extra heavy scalar particles with hypercharge that couple to the hidden sector fields
non-universally, $N_S$ would not vanish.}
In the case that $D$-term contributions are independent of the generation,
the sum rules (\ref{12}), (\ref{13-1}) and (\ref{13-2}) still hold in their presence.

\subsection{$SU(5)$ type}

We consider the case with $SU(5)$ symmetry in the hidden-visible couplings.
In this case, the following relations hold,
\begin{eqnarray}
\hspace{-10mm} &~& N_{\tilde{q}_{1}} = N_{\tilde{u}^*_{R}} = N_{\tilde{e}^*_{R}}, ~~
N_{\tilde{d}^*_{R}} = N_{\tilde{l}_{1}} , ~~
N_{\tilde{q}_{2}} = N_{\tilde{c}^*_{R}} = N_{\tilde{\mu}^*_{R}}, ~~
N_{\tilde{s}^*_{R}} = N_{\tilde{l}_{2}} ,
\nonumber \\ 
\hspace{-10mm} &~& N_{\tilde{q}_{3}} = N_{\tilde{t}^*_{R}} = N_{\tilde{\tau}^*_{R}}, ~~
N_{\tilde{b}^*_{R}} = N_{\tilde{l}_{3}} .
\label{N-SU5}
\end{eqnarray}
Using these relations (\ref{N-SU5}), we derive the following three kinds of sum rules
\begin{eqnarray}
&~& M_{\tilde{u}_{L}}^{2}-M_{\tilde{c}_{L}}^{2} = M_{\tilde{u}_{R}}^{2}-M_{\tilde{c}_{R}}^{2} 
= M_{\tilde{e}_{R}}^{2}-M_{\tilde{\mu}_{R}}^{2} , ~~
\label{SU5-1} \\
&~& M_{\tilde{d}_{R}}^{2}-M_{\tilde{s}_{R}}^{2} = M_{\tilde{e}_{L}}^{2}-M_{\tilde{\mu}_{L}}^{2} .
\label{SU5-2}
\end{eqnarray}
If $N_S = 0$, (\ref{IG}) holds.

\subsection{$SO(10)$ type}

We consider the case with $SO(10)$ symmetry in the hidden-visible couplings.
In this case, the following relations hold,
\begin{eqnarray}
\hspace{-12mm} &~& N_{\tilde{q}_{1}} = N_{\tilde{u}^*_{R}} = N_{\tilde{e}^*_{R}} =
N_{\tilde{d}^*_{R}} = N_{\tilde{l}_{1}} , ~~
N_{\tilde{q}_{2}} = N_{\tilde{c}^*_{R}} = N_{\tilde{\mu}^*_{R}} =
N_{\tilde{s}^*_{R}} = N_{\tilde{l}_{2}} ,
\nonumber \\ 
\hspace{-12mm} &~& N_{\tilde{q}_{3}} = N_{\tilde{t}^*_{R}} = N_{\tilde{\tau}^*_{R}} =
N_{\tilde{b}^*_{R}} = N_{\tilde{l}_{3}} , ~~
N_{h_1} = N_{h_2} .
\label{N-SO10}
\end{eqnarray}
Using these relations (\ref{N-SO10}), we derive (\ref{IG}) and (\ref{13-1}) 
and the following four kinds of sum rules
\begin{eqnarray}
M_{\tilde{u}_{L}}^{2}-M_{\tilde{c}_{L}}^{2} = M_{\tilde{u}_{R}}^{2}-M_{\tilde{c}_{R}}^{2} 
= M_{\tilde{e}_{R}}^{2}-M_{\tilde{\mu}_{R}}^{2} = M_{\tilde{d}_{R}}^{2}-M_{\tilde{s}_{R}}^{2} 
= M_{\tilde{e}_{L}}^{2}-M_{\tilde{\mu}_{L}}^{2} .
\label{SO10}
\end{eqnarray}
In the presence of $D$-term contribution related to $SO(10)/SU(5)$ generator,
the above sum rules (\ref{13-1}) and (\ref{SO10}) still hold on.
A similar feature holds on for the following partially unified types.

\subsection{$SU(5) \times U(1)_{\rm F}$ type}

We consider the case with a flipped $SU(5)$ symmetry in the hidden-visible couplings.
In this case, the following relations hold,
\begin{eqnarray}
\hspace{-10mm} &~& N_{\tilde{q}_{1}} = N_{\tilde{d}^*_{R}} , ~~ N_{\tilde{u}^*_{R}} = N_{\tilde{l}_{1}} , ~~
N_{\tilde{q}_{2}} = N_{\tilde{s}^*_{R}} , ~~ N_{\tilde{c}^*_{R}} = N_{\tilde{l}_{2}} , 
\nonumber \\ 
\hspace{-10mm} &~& N_{\tilde{q}_{3}} = N_{\tilde{b}^*_{R}} , ~~ N_{\tilde{t}^*_{R}} = N_{\tilde{l}_{3}} .
\label{N-F}
\end{eqnarray}
Using these relations (\ref{N-F}), 
we derive the following two kinds of sum rules
\begin{eqnarray}
M_{\tilde{u}_{L}}^{2}-M_{\tilde{c}_{L}}^{2} = M_{\tilde{d}_{R}}^{2}-M_{\tilde{s}_{R}}^{2} , ~~
M_{\tilde{u}_{R}}^{2}-M_{\tilde{c}_{R}}^{2} = M_{\tilde{e}_{L}}^{2}-M_{\tilde{\mu}_{L}}^{2} .
\label{SU5F}
\end{eqnarray}

\subsection{$SU(4) \times SU(2)_L \times SU(2)_R$ type}

We consider the case with $SU(4) \times SU(2)_L \times SU(2)_R$ symmetry in the hidden-visible couplings.
In this case, the following relations hold,
\begin{eqnarray}
\hspace{-10mm} &~& N_{\tilde{q}_{1}} = N_{\tilde{l}_{1}} , ~~ 
N_{\tilde{u}^*_{R}} = N_{\tilde{d}^*_{R}} = N_{\tilde{e}^*_{R}} , ~~
N_{\tilde{q}_{2}} = N_{\tilde{l}_{2}} , ~~ N_{\tilde{c}^*_{R}} = N_{\tilde{s}^*_{R}} = N_{\tilde{\mu}^*_{R}} ,
\nonumber \\ 
\hspace{-10mm} &~& N_{\tilde{q}_{3}} = N_{\tilde{l}_{3}} , ~~ 
N_{\tilde{t}^*_{R}} = N_{\tilde{b}^*_{R}} = N_{\tilde{\tau}^*_{R}} , ~~
N_{h_1} = N_{h_2} .
\label{N-PS}
\end{eqnarray}
Using these relations (\ref{N-PS}), we derive the following three kinds of sum rules
\begin{eqnarray}
M_{\tilde{u}_{L}}^{2}-M_{\tilde{c}_{L}}^{2} = M_{\tilde{e}_{L}}^{2}-M_{\tilde{\mu}_{L}}^{2} , ~~
M_{\tilde{u}_{R}}^{2}-M_{\tilde{c}_{R}}^{2} = M_{\tilde{d}_{R}}^{2}-M_{\tilde{s}_{R}}^{2} 
= M_{\tilde{e}_{R}}^{2}-M_{\tilde{\mu}_{R}}^{2} .
\label{PS}
\end{eqnarray}
If $N_S = 0$, (\ref{IG}) holds.

\subsection{$SU(3)_C \times SU(3)_L \times SU(3)_R$ type}

We consider the case with $SU(3)_C \times SU(3)_L \times SU(3)_R$ symmetry in the hidden-visible couplings.
For sfermions in the first generation, $\tilde{q}_1$ belongs to $({\bf 3}, {\bf 3}, {\bf 1})$,
$\tilde{u}^*_R$ and $\tilde{d}^*_R$ belong to $(\overline{\bf 3}, {\bf 1}, \overline{\bf 3})$ 
and $\tilde{l}_L$ and $\tilde{e}^*_R$ belong to $({\bf 1}, \overline{\bf 3}, {\bf 3})$ of 
$SU(3)_C \times SU(3)_L \times SU(3)_R$.
The same assignment holds on for other generations.
In this case, the following relations hold,
\begin{eqnarray}
\hspace{-10mm} &~& N_{\tilde{u}^*_{R}} = N_{\tilde{d}^*_{R}} , ~~ N_{\tilde{e}^*_{R}} = N_{\tilde{l}_{1}} , ~~
N_{\tilde{c}^*_{R}} = N_{\tilde{s}^*_{R}} , ~~ N_{\tilde{\mu}^*_{R}} = N_{\tilde{l}_{2}} ,
\nonumber \\ 
\hspace{-10mm} &~& N_{\tilde{t}^*_{R}} = N_{\tilde{b}^*_{R}} , ~~ N_{\tilde{\tau}^*_{R}} = N_{\tilde{l}_{3}} .
\label{N-Tri}
\end{eqnarray}
Using these relations (\ref{N-Tri}), 
we derive the following two kinds of sum rules
\begin{eqnarray}
M_{\tilde{u}_{R}}^{2}-M_{\tilde{c}_{R}}^{2} = M_{\tilde{d}_{R}}^{2}-M_{\tilde{s}_{R}}^{2} , ~~
M_{\tilde{e}_{L}}^{2}-M_{\tilde{\mu}_{L}}^{2} = M_{\tilde{e}_{R}}^{2}-M_{\tilde{\mu}_{R}}^{2} .
\label{Tri}
\end{eqnarray}

\section{Sfermion sum rules in orbifold family unification}

We study sfermion sum rules in orbifold family unification models.
Here the orbifold family unification models
are refered as those derived from 
$SU(N)$ gauge theory on $M^4 \times (S^1/Z_2)$, with the gauge symmetry breaking pattern
$SU(N) \to SU(3) \times SU(2) \times SU(r) \times SU(s) \times U(1)^n$,
which is realized with the $Z_2$ parity assignment
\begin{eqnarray}
&~& P_0 = \mbox{diag}(+1, +1, +1, +1, +1, -1, \dots, -1, -1, \dots, -1) , 
\label{P0-SM} \\
&~& P_1 = \mbox{diag}(+1, +1, +1, -1, -1, \underbrace{+1, \dots, +1}_{r}, \underbrace{-1, \dots, -1}_{s}) ,
\label{P1-SM}
\end{eqnarray}
where $s = N-5-r$ and $N \ge 6$ \cite{KK&O}.\footnote{
In the absence of hidden dynamics,
sfermion mass relations and sum rules were studied in this framework \cite{KK,KK2}.
Sfermion masses have been studied from the viewpoint of flavor symmetry and
its violation in $SU(5)$ SUSY orbifold GUT \cite{H&N2}.}
The matrices $P_0$ and $P_1$ are the representation matrices (up to sign factors) of the fundamental representation 
of the $Z_2$ transformation ($y \to -y$) and the $Z_2'$ transformation ($y \to 2\pi R- y$), respectively.
Here, $y$ is the coordinate of $S^1/Z_2$, and $R$ is the radius of $S^1$.
After the breakdown of $SU(N)$, the rank-$k$ completely antisymmetric tensor representation $[N, k]$, 
whose dimension is ${}_{N}C_{k}$, is decomposed into
a sum of multiplets of the subgroup $SU(3) \times SU(2) \times SU(r) \times SU(s)$ as
\begin{eqnarray}
[N, k] = \sum_{l_1 =0}^{k} \sum_{l_2 = 0}^{k-l_1} \sum_{l_3 = 0}^{k-l_1-l_2}  
\left({}_{3}C_{l_1}, {}_{2}C_{l_2}, {}_{r}C_{l_3}, {}_{s}C_{l_4}\right) ,
\label{Nk}
\end{eqnarray}
where $l_1$, $l_2$ and $l_3$ are integers, we have the relation $l_4=k-l_1-l_2-l_3$,
and our notation is such that ${}_{n}C_{l}= 0$ for $l > n$ and $l < 0$.
We define the $Z_2$ parity for the
representation~$({}_{p}C_{l_1}, {}_{q}C_{l_2},$ ${}_{r}C_{l_3},$ ${}_{s}C_{l_4})$ as
\begin{eqnarray}
\mathcal{P}_0 = (-1)^{l_1+l_2} (-1)^k \eta_k ,~~ 
\mathcal{P}_1 = (-1)^{l_1+l_3} (-1)^k \eta'_k ,
\label{Z2}
\end{eqnarray}
where $\eta_k$ and $\eta'_k$ are the intrinsic $Z_2$ parities and
each takes the value $+1$ or $-1$ by definition.
We find that all zero modes of mirror particles are eliminated when we take $(-1)^k \eta_{k} = +1$.
Hereafter, we consider such a case.

We write the flavor numbers of $(d_{R})^c$, $l_{L}$, $(u_{R})^c$, $(e_{R})^c$ and $q_{L}$ 
as $n_{\bar{d}}$, $n_l$, $n_{\bar{u}}$, $n_{\bar{e}}$ and $n_q$.
Both left-handed and right-handed Weyl fermions having even $Z_2$ parities, $\mathcal{P}_0 = \mathcal{P}_1 = +1$, 
compose chiral fermions in the SM.
We list the flavor number of each chiral fermion derived from $[N, k]$
in Table \ref{t1} and \ref{t2}.
\begin{table}[htb]
\caption{The flavor number of each chiral fermion with $(-1)^k \eta_{k} = (-1)^k \eta'_{k} = +1$
and sum rules.}
\label{t1}
\begin{center}
\begin{tabular}{c|c|c|c|c|c|c|l}
\hline
rep. &$(r,s)$&$n_{\bar{d}}$&$n_{l}$&$n_{\bar{u}}$&$n_{\bar{e}}$&$n_{q}$& Sum rules \\
\hline
$[6,3]$ &(0,1)&0&0&2&2&0& $M_{\tilde{u}_R}^2 -M_{\tilde{c}_R}^2 = M_{\tilde{e}_R}^2 -M_{\tilde{\mu}_R}^2 = 0$   \\
\hline
&(2,0)&1&0&1&1&2& $M_{\tilde{u}_L}^2 -M_{\tilde{c}_L}^2 = 0$ \\
\cline{2-8}
$[7,3]$&(1,1)&0&1&2&2&1& $M_{\tilde{u}_R}^2 -M_{\tilde{c}_R}^2 = M_{\tilde{e}_R}^2 -M_{\tilde{\mu}_R}^2 = 0$  \\
\cline{2-8}
&(0,2)&1&0&3&3&0& $M_{\tilde{u}_R}^2 -M_{\tilde{c}_R}^2 = M_{\tilde{e}_R}^2 -M_{\tilde{\mu}_R}^2 = 0$   \\
\hline
&(3,0)&3&0&1&1&3& $M_{\tilde{d}_R}^2 -M_{\tilde{s}_R}^2 = M_{\tilde{u}_L}^2 -M_{\tilde{c}_L}^2 = 0$   \\
\cline{2-8}
$[8,3]$&(2,1)&1&2&2&2&2& $M_{\tilde{u}_R}^2 -M_{\tilde{c}_R}^2 = M_{\tilde{e}_R}^2 -M_{\tilde{\mu}_R}^2$ \\
& & & & & & & $= M_{\tilde{e}_L}^2 -M_{\tilde{\mu}_L}^2 = M_{\tilde{u}_L}^2 -M_{\tilde{c}_L}^2 = 0$ \\
\cline{2-8}
&(1,2)&1&2&3&3&1& $M_{\tilde{u}_R}^2 -M_{\tilde{c}_R}^2 = M_{\tilde{e}_R}^2 -M_{\tilde{\mu}_R}^2 $ \\
& & & & & & & $= M_{\tilde{e}_L}^2 -M_{\tilde{\mu}_L}^2 = 0$ \\
\cline{2-8}
&(0,3)&3&0&4&4&0& $M_{\tilde{u}_R}^2 -M_{\tilde{c}_R}^2 = M_{\tilde{e}_R}^2 -M_{\tilde{\mu}_R}^2$ \\
& & & & & & & $= M_{\tilde{d}_R}^2 -M_{\tilde{s}_R}^2 = 0$ \\
\hline
&(3,0)&1&1&3&3&3& $M_{\tilde{u}_R}^2 -M_{\tilde{c}_R}^2 = M_{\tilde{e}_R}^2 -M_{\tilde{\mu}_R}^2$ \\
& & & & & & & $= M_{\tilde{u}_L}^2 -M_{\tilde{c}_L}^2 = 0$ \\
\cline{2-8}
$[8,4]$&(2,1)&2&0&2&2&4& $M_{\tilde{u}_R}^2 -M_{\tilde{c}_R}^2 = M_{\tilde{e}_R}^2 -M_{\tilde{\mu}_R}^2$ \\
& & & & & & & $= M_{\tilde{d}_R}^2 -M_{\tilde{s}_R}^2 = M_{\tilde{u}_L}^2 -M_{\tilde{c}_L}^2 =0$ \\
\cline{2-8}
&(1,2)&1&1&3&3&3& $M_{\tilde{u}_R}^2 -M_{\tilde{c}_R}^2 = M_{\tilde{e}_R}^2 -M_{\tilde{\mu}_R}^2$ \\
& & & & & & & $= M_{\tilde{u}_L}^2 -M_{\tilde{c}_L}^2 =0$ \\
\cline{2-8}
&(0,3)&2&0&6&6&0& $M_{\tilde{u}_R}^2 -M_{\tilde{c}_R}^2 = M_{\tilde{e}_R}^2 -M_{\tilde{\mu}_R}^2$ \\
& & & & & & & $= M_{\tilde{d}_R}^2 -M_{\tilde{s}_R}^2 =0$ \\
\hline
&(4,0)&6&0&1&1&4& $M_{\tilde{u}_L}^2 -M_{\tilde{c}_L}^2 = M_{\tilde{d}_R}^2 -M_{\tilde{s}_R}^2 =0$ \\
\cline{2-8}
&(3,1)&3&3&2&2&3& $M_{\tilde{u}_R}^2 -M_{\tilde{c}_R}^2 = M_{\tilde{e}_R}^2 -M_{\tilde{\mu}_R}^2$ \\
& & & & & & & $= M_{\tilde{d}_R}^2 -M_{\tilde{s}_R}^2 = M_{\tilde{e}_L}^2 -M_{\tilde{\mu}_L}^2$ \\
& & & & & & & $= M_{\tilde{u}_L}^2 -M_{\tilde{c}_L}^2 =0$ \\
\cline{2-8}
$[9,3]$&(2,2)&2&4&3&3&2& $M_{\tilde{u}_R}^2 -M_{\tilde{c}_R}^2 = M_{\tilde{e}_R}^2 -M_{\tilde{\mu}_R}^2$ \\
& & & & & & & $= M_{\tilde{d}_R}^2 -M_{\tilde{s}_R}^2 = M_{\tilde{e}_L}^2 -M_{\tilde{\mu}_L}^2=0$, \\
& & & & & & & $2\left(M_{\tilde{e}_{L}}^{2}-M_{\tilde{\tau}_{L}}^{2}\right) 
= M_{\tilde{e}_{R}}^{2} - M_{\tilde{\tau}_{R}}^{2}$ \\
\cline{2-8}
&(1,3)&3&3&4&4&1& $M_{\tilde{u}_R}^2 -M_{\tilde{c}_R}^2 = M_{\tilde{e}_R}^2 -M_{\tilde{\mu}_R}^2$ \\
& & & & & & & $= M_{\tilde{d}_R}^2 -M_{\tilde{s}_R}^2 = M_{\tilde{e}_L}^2 -M_{\tilde{\mu}_L}^2 =0$, \\
& & & & & & & $2\left(M_{\tilde{e}_{L}}^{2}-M_{\tilde{\tau}_{L}}^{2}\right) 
= M_{\tilde{e}_{R}}^{2} - M_{\tilde{\tau}_{R}}^{2}$ \\
\cline{2-8}
&(0,4)&6&0&5&5&0& $M_{\tilde{u}_R}^2 -M_{\tilde{c}_R}^2 = M_{\tilde{e}_R}^2 -M_{\tilde{\mu}_R}^2$ \\
& & & & & & & $=M_{\tilde{d}_R}^2 -M_{\tilde{s}_R}^2 =0$ \\
\hline
\end{tabular}
\end{center}
\end{table}

\begin{table}[htb]
\caption{The flavor number of each chiral fermion with $(-1)^k \eta_{k} = +1, (-1)^k \eta'_{k} = -1$
and sum rules.}
\label{t2}
\begin{center}
\begin{tabular}{c|c|c|c|c|c|c|l}
\hline
rep. &$(r,s)$&$n_{\bar{d}}$&$n_{l}$&$n_{\bar{u}}$&$n_{\bar{e}}$&$n_{q}$& Sum rules \\
\hline
$[6,3]$&(0,1)&0&0&0&0&2& $M_{\tilde{u}_L}^2 -M_{\tilde{c}_L}^2 = 0$ \\
\hline
&(2,0)&0&1&2&2&1& $M_{\tilde{u}_R}^2 -M_{\tilde{c}_R}^2 = M_{\tilde{e}_R}^2 -M_{\tilde{\mu}_R}^2 = 0$ \\
\cline{2-8}
$[7,3]$&(1,1)&1&0&1&1&2& $M_{\tilde{u}_L}^2 -M_{\tilde{c}_L}^2 = 0$ \\
\cline{2-8}
&(0,2)&0&1&0&0&3& $M_{\tilde{u}_L}^2 -M_{\tilde{c}_L}^2 = 0$ \\
\hline
&(3,0)&0&3&3&3&1& $M_{\tilde{u}_R}^2 -M_{\tilde{c}_R}^2 = M_{\tilde{e}_R}^2 -M_{\tilde{\mu}_R}^2$ \\
& & & & & & & $= M_{\tilde{e}_L}^2 -M_{\tilde{\mu}_L}^2 = 0$, \\
& & & & & & & $2\left(M_{\tilde{e}_{L}}^{2}-M_{\tilde{\tau}_{L}}^{2}\right) 
= M_{\tilde{e}_{R}}^{2} - M_{\tilde{\tau}_{R}}^{2}$ \\
\cline{2-8}
$[8,3]$&(2,1)&2&1&2&2&2& $M_{\tilde{u}_R}^2 -M_{\tilde{c}_R}^2 = M_{\tilde{e}_R}^2 -M_{\tilde{\mu}_R}^2$ \\
& & & & & & & $= M_{\tilde{d}_R}^2 -M_{\tilde{s}_R}^2 = M_{\tilde{u}_L}^2 -M_{\tilde{c}_L}^2 = 0$ \\
\cline{2-8}
&(1,2)&2&1&1&1&3& $M_{\tilde{d}_R}^2 -M_{\tilde{s}_R}^2 = M_{\tilde{u}_L}^2 -M_{\tilde{c}_L}^2 = 0$ \\
\cline{2-8}
&(0,3)&0&3&0&0&4& $M_{\tilde{e}_L}^2 -M_{\tilde{\mu}_L}^2 = M_{\tilde{u}_L}^2 -M_{\tilde{c}_L}^2 = 0$ \\
\hline
&(3,0)&1&1&3&3&3& $M_{\tilde{u}_R}^2 -M_{\tilde{c}_R}^2 = M_{\tilde{e}_R}^2 -M_{\tilde{\mu}_R}^2$ \\
& & & & & & & $= M_{\tilde{u}_L}^2 -M_{\tilde{c}_L}^2 = 0$ \\
\cline{2-8}
$[8,4]$&(2,1)&0&2&4&4&2& $M_{\tilde{u}_R}^2 -M_{\tilde{c}_R}^2 = M_{\tilde{e}_R}^2 -M_{\tilde{\mu}_R}^2$ \\
& & & & & & & $= M_{\tilde{e}_L}^2 -M_{\tilde{\mu}_L}^2 = M_{\tilde{u}_L}^2 -M_{\tilde{c}_L}^2 = 0$ \\
\cline{2-8}
&(1,2)&1&1&3&3&3& $M_{\tilde{u}_R}^2 -M_{\tilde{c}_R}^2 = M_{\tilde{e}_R}^2 -M_{\tilde{\mu}_R}^2$ \\
& & & & & & & $= M_{\tilde{u}_L}^2 -M_{\tilde{c}_L}^2 = 0$ \\
\cline{2-8}
&(0,3)&0&2&0&0&6& $M_{\tilde{e}_L}^2 -M_{\tilde{\mu}_L}^2 = M_{\tilde{u}_L}^2 -M_{\tilde{c}_L}^2 = 0$ \\ 
\hline
&(4,0)&0&6&4&4&1& $M_{\tilde{u}_R}^2 -M_{\tilde{c}_R}^2 = M_{\tilde{e}_R}^2 -M_{\tilde{\mu}_R}^2$ \\
& & & & & & & $= M_{\tilde{e}_L}^2 -M_{\tilde{\mu}_L}^2 = 0$, \\
& & & & & & & $2\left(M_{\tilde{e}_{L}}^{2}-M_{\tilde{\tau}_{L}}^{2}\right) 
= M_{\tilde{e}_{R}}^{2} - M_{\tilde{\tau}_{R}}^{2}$ \\
\cline{2-8}
&(3,1)&3&3&3&3&2& $M_{\tilde{u}_R}^2 -M_{\tilde{c}_R}^2 = M_{\tilde{e}_R}^2 -M_{\tilde{\mu}_R}^2$ \\
& & & & & & & $= M_{\tilde{d}_R}^2 -M_{\tilde{s}_R}^2 = M_{\tilde{e}_L}^2 -M_{\tilde{\mu}_L}^2$ \\
& & & & & & & $= M_{\tilde{u}_L}^2 -M_{\tilde{c}_L}^2 = 0$, \\
& & & & & & & $2\left(M_{\tilde{e}_{L}}^{2}-M_{\tilde{\tau}_{L}}^{2}\right) 
= M_{\tilde{e}_{R}}^{2} - M_{\tilde{\tau}_{R}}^{2}$ \\
\cline{2-8}
$[9,3]$&(2,2)&4&2&2&2&3& $M_{\tilde{u}_R}^2 -M_{\tilde{c}_R}^2 = M_{\tilde{e}_R}^2 -M_{\tilde{\mu}_R}^2$ \\
& & & & & & & $= M_{\tilde{d}_R}^2 -M_{\tilde{s}_R}^2 = M_{\tilde{e}_L}^2 -M_{\tilde{\mu}_L}^2$ \\
& & & & & & & $= M_{\tilde{u}_L}^2 -M_{\tilde{c}_L}^2 = 0$ \\
\cline{2-8}
&(1,3)&3&3&1&1&4& $M_{\tilde{d}_R}^2 -M_{\tilde{s}_R}^2 = M_{\tilde{e}_L}^2 -M_{\tilde{\mu}_L}^2$ \\
& & & & & & & $= M_{\tilde{u}_L}^2 -M_{\tilde{c}_L}^2 = 0$ \\
\cline{2-8}
&(0,4)&0&6&0&0&5& $M_{\tilde{e}_L}^2 -M_{\tilde{\mu}_L}^2 = M_{\tilde{u}_L}^2 -M_{\tilde{c}_L}^2 = 0$ \\
\hline
\end{tabular}
\end{center}
\end{table}


We add the following assumptions in our analysis.\\
1. Three families in the MSSM come from zero modes of the bulk field 
with the representation $[N, k]$ and some brane fields.
Higgs fields originate from other multiplets.
Chiral anomalies may arise at the boundaries with the appearance of chiral fermions.
Such anomalies must be canceled in the four-dimensional effective theory by the contribution of the brane chiral fermions
and/or counterterms, such as the Chern-Simons term \cite{C&H,AC&G,KK&L}.\\
2. We do not specify the mechanism by which the $N=1$ SUSY is broken in four dimensions.\footnote{
The Scherk-Schwarz mechanism, in which SUSY is broken by the difference between the BCs of
bosons and fermions, is typical \cite{S&S,S&S2}.
This mechanism on $S^1/Z_2$ leads to a restricted type of soft SUSY breaking parameters, such as
$M_i = \beta/R$ for bulk gauginos and $m_{\tilde{F}}^2 = (\beta/R)^2$ for bulk scalar particles,
where $\beta$ is a real parameter and $R$ is the radius of $S^1$.}
Soft SUSY breaking terms respect the gauge invariance.\\
3. Extra gauge symmetries are broken by the Higgs mechanism simultaneously 
with the orbifold breaking at the scale $M = O(1/R)$.
Then there can appear extra contributions to soft SUSY breaking parameters.
We need to specify the particle assignment and interactions in order to consider
such contributions.
We do not consider them for simplicity.\\
4. Chiral fermions are first and/or second generation ones 
in the case that the flavor number of each chiral fermion is less than three.\\

Under the above assumptions, specific sum rules among sfermion masses are derived and listed 
in 8-th column of Table 1 and 2.

\section{Conclusions}

We have derived sum rules among scalar masses in the presence of hidden sector dynamics
and shown that they still can be useful probes of the MSSM and beyond.
Scalar sum rules can be classified into following types.

~~\\
(Type A) The EWS sum rules:
\begin{eqnarray}
&~& M_{\tilde{u}_L}^2 - M_{\tilde{d}_L}^2 = M_{\tilde{\nu}_{eL}}^2 - M_{\tilde{e}_L}^2 = M_{\tilde{c}_L}^2 - M_{\tilde{s}_L}^2 
= M_{\tilde{\nu}_{\mu L}}^2 - M_{\tilde{\mu}_L}^2 
\nonumber \\
&~& ~~~ = M_{\tilde{t}_L}^2 - M_{\tilde{b}_L}^2 - m_t^2 =
 M_{\tilde{\nu}_{\tau L}}^2 - M_{\tilde{\tau}_L}^2 = M_W^2 \cos2\beta .
\label{I} 
\end{eqnarray}
These sum rules are derived from the fact that left-handed fermions (and its superpartner) form $SU(2)_L$ doublets,
and they are irrelevant to the structure of models beyond the MSSM.
The sfermion sector (and the breakdown of electroweak symmetry) in the MSSM can be tested by using them.

~~\\
(Type B) Intrafamily sfermion sum rule:
\begin{eqnarray}
2M_{\tilde{u}_{R}}^{2}-M_{\tilde{d}_{R}}^{2}-M_{\tilde{d}_{L}}^{2}+M_{\tilde{e}_{L}}^{2}-M_{\tilde{e}_{R}}^{2} 
 = \frac{10}{3}\left(M_{Z}^{2}-M_{W}^{2}\right)\cos 2\beta .
\label{IF}
\end{eqnarray}
In the case with $N_S = 0$, the universality in each family can be checked by using it.

~~\\
(Type C) Outer-family sfermion sum rules:
\begin{eqnarray}
&~& M_{\tilde{u}_{L}}^{2}-M_{\tilde{c}_{L}}^{2} = M_{\tilde{u}_{R}}^{2}-M_{\tilde{c}_{R}}^{2} 
= M_{\tilde{d}_{R}}^{2}-M_{\tilde{s}_{R}}^{2} 
\nonumber \\
&~& ~~~~~ = M_{\tilde{e}_{L}}^{2}-M_{\tilde{\mu}_{L}}^{2} = M_{\tilde{e}_{R}}^{2}-M_{\tilde{\mu}_{R}}^{2} ,
\label{12f}\\
&~& 2\left(M_{\tilde{u}_{L}}^{2}-M_{\tilde{t}_{L}}^{2} + m_t^2\right) 
= M_{\tilde{u}_{R}}^{2} + M_{\tilde{d}_{R}}^{2} - M_{\tilde{t}_{R}}^{2} - M_{\tilde{b}_{R}}^{2}  + m_t^2 ,
\label{13f1}\\
&~& 2\left(M_{\tilde{e}_{L}}^{2}-M_{\tilde{\tau}_{L}}^{2}\right) 
= M_{\tilde{e}_{R}}^{2} - M_{\tilde{\tau}_{R}}^{2} .
\label{13f2}
\end{eqnarray}
Some of these sum rules are derived from the case that some chiral multiplets form a member of multiple
under some large gauge group.
Hence the sfermion sector with the grand unification can be tested and
the gauge group can be specified by using them.

~~\\
(Type D) $Z_2$ orbifold sfermion sum rules:
\begin{eqnarray}
 M_{\tilde{u}_{R}}^{2} - M_{\tilde{c}_{R}}^{2} = M_{\tilde{e}_{R}}^{2} - M_{\tilde{\mu}_{R}}^{2} .
\label{OFU} 
\end{eqnarray}
This sum rule is a piece of type C and
it is derived on the orbifold breaking of $SU(N)$ gauge symmetry for bulk fields with 
an antisymmetric representation if the bulk field contains ${\bf 10}_L$ or ${\overline{\bf 10}}_R$ 
under the subgroup $SU(5)$, and $SU(2)_L$ singlets have even $Z_2$ parities
in the five-dimensional orbifold grand unification.
The $Z_2$ orbifold breaking (of $SU(N)$ gauge symmetry) can be tested by using it.\\

It is known that the dangerous FCNC processes can be avoided if the sfermion masses
in the first two families are degenerate or rather heavy or fermion and its superpartner mass matrices are aligned.
We have derived sfermion sum rules without a requirement of the mass degeneracy for each squark and slepton species 
in the first two generations.
If we require the mass degeneracy, we obtain the following relations, in most GUTs,
\begin{eqnarray}
&~&  M_{\tilde{u}_{L}}^{2} = M_{\tilde{c}_{L}}^{2} , ~~ M_{\tilde{u}_{R}}^{2} = M_{\tilde{c}_{R}}^{2} , ~~
 M_{\tilde{e}_{R}}^{2} = M_{\tilde{\mu}_{R}}^{2} , 
\nonumber \\
&~&  M_{\tilde{d}_{R}}^{2} = M_{\tilde{s}_{R}}^{2} , ~~ M_{\tilde{e}_{L}}^{2} = M_{\tilde{\mu}_{L}}^{2} .
\label{FCNC} 
\end{eqnarray}
In this case, sum rules including third generation sfermions could be useful to specify models.

In the case that the gauge mediation is dominant, 
the couplings $k_{\tilde{F}}^{(v)}$ parametrize as $k_{\tilde{F}}^{(v)}(0) = \sum_i C_2^{(i)}(\tilde{F}) K_i$ 
using SUSY breaking and messenger dependent functions $K_i$.
Hence the following extra sum rule is derived,\cite{CR&S}
\begin{eqnarray}
 3\left(M_{\tilde{d}_{R}}^{2} - M_{\tilde{u}_{R}}^{2}\right) + M_{\tilde{e}_{R}}^{2} 
=  4\left(M_{Z}^{2}-M_{W}^{2}\right)\cos 2\beta ,
\label{gauge-mediation1} 
\end{eqnarray}
as intrafamily sfermion sum rule in addition (\ref{IG}) for universal type.
For outer-family sfermion sum rules, the degeneracy occurs in the first and second generation sfermion masses
and then some of (\ref{FCNC}) are derived.

If the hidden sector dynamics were strong and superconformal, 
conformal sequestering can occur and anomaly mediation can be dominant.\cite{S,IINSY,SS,KMS,MNP}\footnote{
SUSY standard models coupled with superconformal theories were also studied in Refs. \cite{KT,KNT}.}

The scalar mass relations and sum rules have been also derived 
in various models \cite{P&P,P&P2,K&T,Ch&H,K&M,K&K,KK&K,A&P,A&P2,IK&Y}.
In the future, we expect that they can be useful to probe a physics beyond the MSSM.

\section*{Acknowledgements}
This work was supported in part by Scientific Grants from the Ministry of Education and Science, 
Grant No.18204024 and Grant No.18540259 (Y.K.).


\begin{thebibliography}{99}
  \bibitem{N}
     H.~P.~Nilles, Phys.~Rep. {\bf 110}, 1 (1984).
  \bibitem{H&K}
     H.~Haber and G.~L.~Kane, Phys.~Rep. {\bf 117}, 75 (1985).    
  \bibitem{unif}
     C.~Giunti, C.~W.~Kim and U.~W.~Lee, Mod.~Phys.~Lett. {\bf A6}, 1745 (1991).
  \bibitem{unif2}
     J.~Ellis, S.~Kelley and D.~V.~Nanopoulos, Phys.~Lett. {\bf B260}, 131 (1991).
  \bibitem{unif3}
     U.~Amaldi, W.~de~Boer and H.~Furstenau, Phys.~Lett. {\bf B260}, 447 (1991).
  \bibitem{unif4}
     P.~Langacker and M.~Luo, Phys.~Rev. {\bf D44}, 817 (1991).
  \bibitem{FHK&N}
     A.~E.~Faraggi, J.~S.~Hagelin, S.~Kelley and D.~V.~Nanopoulos, Phys.~Rev. {\bf D45}, 3272 (1992).
  \bibitem{M&R}
     S.~P.~Martin and P.~Ramond, Phys.~Rev. {\bf D48}, 5365 (1993).
  \bibitem{KM&Y}
     Y.~Kawamura, H.~Murayama and M.~Yamaguchi, Phys.~Lett. {\bf B324}, 52 (1994).
  \bibitem{KM&Y2}
     Y.~Kawamura, H.~Murayama and M.~Yamaguchi, Phys.~Rev. {\bf D51}, 1337 (1995).
  \bibitem{DFGSS&T}
     M.~Dine, P.~J.~Fox, E.~Gorbatov, Y.~Shadmi, Y.~Shirman and S.~Thomas, Phys.~Rev. {\bf D70}, 045023 (2004).
  \bibitem{S}
     R.~Sundrum, Phys.~Rev.~{\bf D71}, 085003 (2005).
  \bibitem{IINSY}
     M.~Ibe, K.-I.~Izawa, Y.~Nakayama, Y.~Shinbara and T.~Yanagida,
     Phys.~Rev.~{\bf D73}, 015004 (2006): Phys.~Rev.~{\bf D73}, 035012 (2006).
  \bibitem{SS}
     M.~Schmaltz and R.~Sundrum, J. High Energy Phys. {\bf 11}, 011 (2006).
  \bibitem{KMS}
     S.~Kachru, L.~McAllister and R.~Sundrum, J. High Energy Phys. {\bf 10}, 013 (2007).
  \bibitem{MNP}
     H.~Murayama, Y.~Nomura and D.~Poland, Phys.~Rev.~{\bf D77}, 015005 (2008).
  \bibitem{CR&S}
     A.~G.~Cohen, T.~S.~Roy and M.~Schmaltz, J. High Energy Phys. {\bf 02}, 027 (2007).
  \bibitem{Dterm}
     M.~Drees, Phys.~Lett. {\bf B181}, 279 (1986).
  \bibitem{Dterm2}
     J.~S.~Hagelin and S.~Kelley, Nucl.~Phys. {\bf B342}, 95 (1990).
  \bibitem{Kazakov}
    D.~I.~Kazakov, {\it Supersymmetry in particle physics:
    the renormalization group viewpoint}, hep-ph/0001257.
  \bibitem{Polonsky}
    N.~Polonsky, {\it Supersymmetry: Structure and Phenomena - 
    Extensions of the Standard Model}, Lecture Notes in Physics Monograph {\bf 68} (Springer-Verlag, 2001).
  \bibitem{DG&R}
    M.~Drees, R.~M.~Godbole and P.~Roy, {\it Theory and Phenomenology of Sparticles} (World Scientific, 2004).
  \bibitem{B&T}
    H.~Baer and X.~Tata, {\it Weak Scale Supersymmetry} (Cambridge, 2006).
  \bibitem{M&V}
     S.~Martin and M.~Vaughn, Phys.~Lett. {\bf B318}, 331 (1993).
  \bibitem{KK&O}
     Y.~Kawamura, T.~Kinami and K.~Oda, Phys.~Rev. {\bf D76}, 035001 (2007).
  \bibitem{KK}
     Y.~Kawamura and T.~Kinami, Int.~J.~Mod.~Phys. {\bf A22}, 4617 (2007)
  \bibitem{KK2}
     Y.~Kawamura and T.~Kinami, Prog.~Theor.~Phys. {\bf 119}, 285 (2008).
  \bibitem{H&N2}
     L.~Hall and Y.~Nomura, Ann. of Phys. {\bf 306}, 132 (2003).
  \bibitem{C&H}
     C.~G.~Callan and J.~A.~Harvey, Nucl.~Phys. {\bf B250}, 427 (1985).
  \bibitem{AC&G}
     N.~Arkani-Hamed, A.~G.~Cohen and H.~Georgi, Phys.~Lett. {\bf B516}, 395 (2001).
  \bibitem{KK&L}
     H.~D.~Kim, J.~E.~Kim and H.~M.~Lee, J. High Energy Phys. {\bf 06}, 048 (2002).
  \bibitem{S&S}
     J.~Scherk and J.~H.~Schwarz, Phys.~Lett. {\bf B82}, 60 (1979)
  \bibitem{S&S2}
     J.~Scherk and J.~H.~Schwarz, Nucl.~Phys. {\bf B153}, 61 (1979).
  \bibitem{KT}
     T.~Kobayashi and H.~Terao, Phys.~Rev.~{\bf D64}, 075003 (2001).
  \bibitem{KNT}
     T.~Kobayashi and H.~Terao, Phys.~Rev.~{\bf D65}, 015006 (2002).
  \bibitem{P&P}
     N.~Polonsky and A.~Pomarol, Phys.~Rev.~Lett. {\bf 73}, 2292 (1994).
  \bibitem{P&P2}
     N.~Polonsky and A.~Pomarol, Phys.~Rev. {\bf D51}, 6532 (1995).
  \bibitem{K&T}
     Y.~Kawamura and M.~Tanaka, Prog.~Theor.~Phys. {\bf 91}, 949 (1994).
  \bibitem{Ch&H}
     H.~C.~Cheng and L.~J.~Hall, Phys.~Rev. {\bf D51}, 5289 (1995).
  \bibitem{K&M}
     C.~Kolda and S.~P.~Martin, Phys.~Rev. {\bf D53}, 3871 (1996).
  \bibitem{KK&K}
     Y.~Kawamura, T.~Kobayashi and T.~Komatsu, Phys.~Lett. {\bf B400}, 284 (1997).
  \bibitem{K&K}
     Y.~Kawamura and T.~Kobayashi, Phys.~Rev. {\bf D56}, 3844 (1997).
  \bibitem{A&P}
     B.~Ananthanarayan and P.~N.~Pandita, Mod.~Phys.~Lett. {\bf A19}, 467 (2004).
  \bibitem{A&P2}
     B.~Ananthanarayan and P.~N.~Pandita, Int. J. Mod. Phys. {\bf A20}, 4241 (2005);
     Int. J. Mod. Phys. {\bf A22}, 3229 (2007).
  \bibitem{IK&Y}
     K.~Inoue, K.~Kojima and K.~Yoshioka, J. High Energy Phys. {\bf 07}, 027 (2007).

\end{thebibliography}
\end{document}